\documentclass[twocolumn,amsmath,amssymb,floatfix,nature,shownopacs,footinbib,superscriptaddress]{revtex4-1}
\usepackage{amsmath,amssymb,natbib,bm,url,epsf,color,siunitx,float}

\usepackage[pdftex]{graphicx}
\usepackage{xcolor}
\usepackage{dcolumn}
\usepackage{cancel}
\usepackage{comment}

\usepackage{url}

\usepackage{subcaption}

\begin{document}


\title{Patterns in temporal networks with higher-order egocentric structures}

\author{Beatriz Arregui-Garc\'ia}
\affiliation{Instituto de F\'{\i}sica Interdisciplinar y Sistemas Complejos IFISC (CSIC-UIB), Campus UIB, 07122 Palma de Mallorca, Spain. }
\author{Antonio Longa}
\affiliation{University of Trento, Trento, Italy}
\author{Quintino Francesco Lotito}
\affiliation{University of Trento, Trento, Italy}
\author{Sandro Meloni}
\affiliation{Instituto de F\'{\i}sica Interdisciplinar y Sistemas Complejos IFISC (CSIC-UIB), Campus UIB, 07122 Palma de Mallorca, Spain. }
\author{Giulia Cencetti}\email{giulia.cencetti@cpt.univ-mrs.fr}
\affiliation{Aix-Marseille Univ, Université de Toulon, CNRS, CPT, Marseille, France}

\date{\today}
\begin{abstract}
The analysis of complex and time-evolving interactions like  social dynamics represents a current challenge for the science of complex systems.
Temporal networks stand as a suitable tool to schematise such systems, encoding all the appearing interactions between pairs of individuals in discrete time. 
Over the years, network science has developed many measures to analyse and compare temporal networks. 
Some of them imply a decomposition of the network into small pieces of interactions, i.e. only involving a few nodes for a short time range. 
Along this line, a possible way to decompose a network is to assume an egocentric perspective, i.e. to consider for each node the time evolution of its neighbourhood.
This has been proposed by Longa et al.~\cite{Longa2022} by defining the ``egocentric temporal neighbourhood'', which has proven a useful tool to characterise temporal networks relative to social interactions.
However, this definition neglects group interactions (quite common in social domains) as they are always decomposed into pairwise connections.
A more general framework that allows us to consider also larger interactions is represented by higher-order networks. 
Here, we generalise the description of social interactions by making use of hypergraphs, consequently, we generalise its decomposition into ``hyper egocentric temporal neighbourhoods''.
This will allow us to analyse social interactions, to compare different datasets or different nodes inside a dataset, by taking into account the intrinsic complexity represented by higher-order interactions.
Even if we limit the order of interactions to the second (triplets of nodes), our results reveal the importance of a higher-order representation. In fact, our analyses show second-order structures are responsible for the majority of the variability at all scales: between datasets, amongst nodes and over time. 
\end{abstract}

\maketitle

\section{Introduction}

Networks, characterised by interconnected nodes and edges, have become essential tools for modelling a multitude of phenomena, ranging from social interactions to biological and physical systems~\cite{newman2018networks, posfai2016network, boccaletti2006complex}.
While they have proven to be a valuable framework, they frequently fail to capture the entire complexity of real-world systems. For example, networks can only capture static interactions, while many real-world interactions tend to be dynamic, i.e., they are established and destroyed during the life of a system. These nuances are better captured by more sophisticated frameworks, such as the temporal networks~\cite{holme2012temporal,masuda2016guide}. We use temporal networks for instance to describe the synchronous activation of remote areas of the brain~\cite{bassett2011dynamic}, social networks~\cite{lachi2023impact}, blockchain~\cite{galdeman2022disentangling}, mobility~\cite{simini2021deep}, emails or phone call communications~\cite{pan2011path}, and the rapidly varying physical interactions between individuals in a closed environment~\cite{barrat2013temporal}. 

The flexibility of temporal networks to describe time-varying interactions, however, comes at the cost of a more complex representation and analysis. In particular, it could be hard to identify relevant structures and nodes' behaviour through the time evolution of the system.

To overcome this problem, it could be useful to decompose networks into small pieces~\cite{milo2002network, kovanen2011temporal, paranjape2017motifs, battiston2017multilayer, lee2020hypergraph, lotito2022higher, lotito2023exact}, i.e., a few nodes and their connections for limited temporal snapshots~\cite{kovanen2011temporal, paranjape2017motifs}, and use these repeated patterns to characterise and compare different networks~\cite{milo2004superfamilies}. Along this line, one possible strategy is to assume a so-called \textit{egocentric perspective}, as proposed recently by Longa et al.~\cite{Longa2022}. This approach consists of considering the evolution of the neighbourhood of a node for a small number of temporal layers. These small temporal subgraphs can be collected for every node of a network and at every time step. 
The entire set of Egocentric Temporal Neighbourhood (ETN) can be reduced to a smaller set of only significant structures with respect to a null model,  Egocentric Temporal Motifs (ETM). ETN and ETM allow for the efficient identification of repeated interaction patterns among individuals in social settings. The ego perspective indeed simplifies motif identification by comparing egocentric temporal sub-networks using a signature, a bit vector, thus avoiding the computational complexities associated with standard motif mining (which usually includes the graph isomorphism problem). 

However, a drawback of this approach is that it neglects connections among the neighbours of the ego node in each timestamp, thus destroying correlations in the creation of links between different ego nodes, a feature of paramount importance, for example, in temporal social interactions~\cite{cencetti2021temporal}.

In recent years and to take into account such more complex structures, a new branch of network science has emerged with the study of higher-order networks~\cite{battiston2020networks, battiston2021physics} and their temporal features~\cite{cencetti2021temporal, comrie2021hypergraph, ceria2023temporal, gallo2023higher, iacopini2023temporal, digaetano2024percolation}. 
In this view, mathematical tools like hypergraphs can be used to represent interactions that can involve any number of nodes at the same time, thus, considering correlations in the behaviour of three or more nodes, effectively extending the concept of edges to hyperedges. The number of involved nodes defines the order of the hyperedge: first order for couples of nodes, second order for triads, and so on. Hypergraphs proved to be important to describe, for instance:  scientific collaborations~\cite{lung2018hypergraph}, relations among species in ecosystems~\cite{grilli2017higher,levine2017beyond}, and simultaneous group interactions in social environments~\cite{sahasrabuddhe2021modelling,cencetti2021temporal} among others.

Given these two different views, the node egocentric from one side and the higher-order from the other, the idea of combining them naturally arises. 
We therefore investigate how to characterise temporal networks via higher-order egocentric structures.
In order to do this we generalise the concept of ETN from graphs to hypergraphs, defining the Hyper Egocentric Temporal Neighbourhood (HETN). 
The new framework will allow us to take into account more complex information including hyperedges and their time evolution.
We will analyse 10 datasets of social interactions. In order to give a thorough representation of such complex and time-evolving systems, we will make use of temporal hypergraphs with hyperedges up to the second order (larger interactions are decomposed into groups of three nodes).
The HETN will encode a description of the temporal hypergraph at the micro-scale with an egocentric perspective. 
We will use them to compare different datasets and different nodes inside a dataset.
Our results make manifest the importance of second order of interactions in describing temporal dynamics as they are responsible for a large part of the variability observed at all scales: between datasets, amongst individuals and, even, for the same individual, over time.

\section{Methods}

\subsection{Hyper Egocentric Temporal Neighbourhood}


We define a \textit{temporal hypergraph} $\mathcal{H} = (V, E)$ as an ensemble comprising a set of nodes $V$ and a set of temporal hyperedges $E$ that encode the higher-order interactions between nodes and the time at which they take place. Time is discrete and represented by natural numbers. For simplicity, we consider hyperedges up to the second order of interactions, i.e. involving three nodes at maximum. The hyperedges of the first order can be represented as triplets $(i, j, t)$, where $i, j \in V$ and $t \in [0, T]$. Hyperedges of the second order are depicted as quadruplets $(i, j, k, t)$, where $i, j, k \in V$. Here, $t$ denotes the time at which the interaction takes place and $T$ corresponds to the time of the last temporal layer in the hypergraph.
\\
A \textit{temporal hypergraph snapshot} is a static graph that represents one temporal layer of $\mathcal{H}$ and includes only interactions taking place at a specific time. It is defined as $\mathcal{H}_t = (V, E_t)$ where $E_t$ is the set of hyperedges $(i, j)$ and $(i, j, k)$ that are associated to time $t$.

We can now define the \textit{hyper egocentric temporal neighbourhood} (HETN), by extending to hypergraphs the concept introduced by \cite{Longa2022} for simple graphs.\\
To do that, we first need to define the \textit{hyper egocentric neighbourhood} of a designated node $v \in V$, i.e. the ego node, as the subgraph $\mathcal{H}_t(v)$ of $\mathcal{H}_t$ comprising all hyperedges that include $v$ at time $t$, i.e. first order edges $(v,j)$ and second order edges $(v, j, k)$. \\
Thus, a HETN of time-length $k$, $\mathcal{H}_{t}^k(v)$, is defined as the sequence of $k$ temporal hypergraph snapshots starting from time $t$: $\{ \mathcal{H}_{t}(v), \mathcal{H}_{t+1}(v), ..., \mathcal{H}_{t+k} (v) \}$. In a HETN, every node is linked to its next occurrence (if any) along the sequence (see Fig.~\ref{fig:fig_1}a). In other words, $\mathcal{H}_{t}^k(v)$ is a temporal subgraph of $\mathcal{H}$ including the higher-order neighbourhood of node $v$ for $k$ temporal layers starting from time $t$. In the rest of the work, following the analyses done in \cite{Longa2022}, we set the time-length of the HETN to $k=2$, i.e. $\mathcal{H}_t^2(v)$, with the aim of reducing the complexity of the signatures and thus having a concise representation of the datasets.

\subsection{Hyper Egocentric Temporal Neighbourhood signatures}

\begin{figure*}[t]
    \centering
    \includegraphics[width=0.9\linewidth]{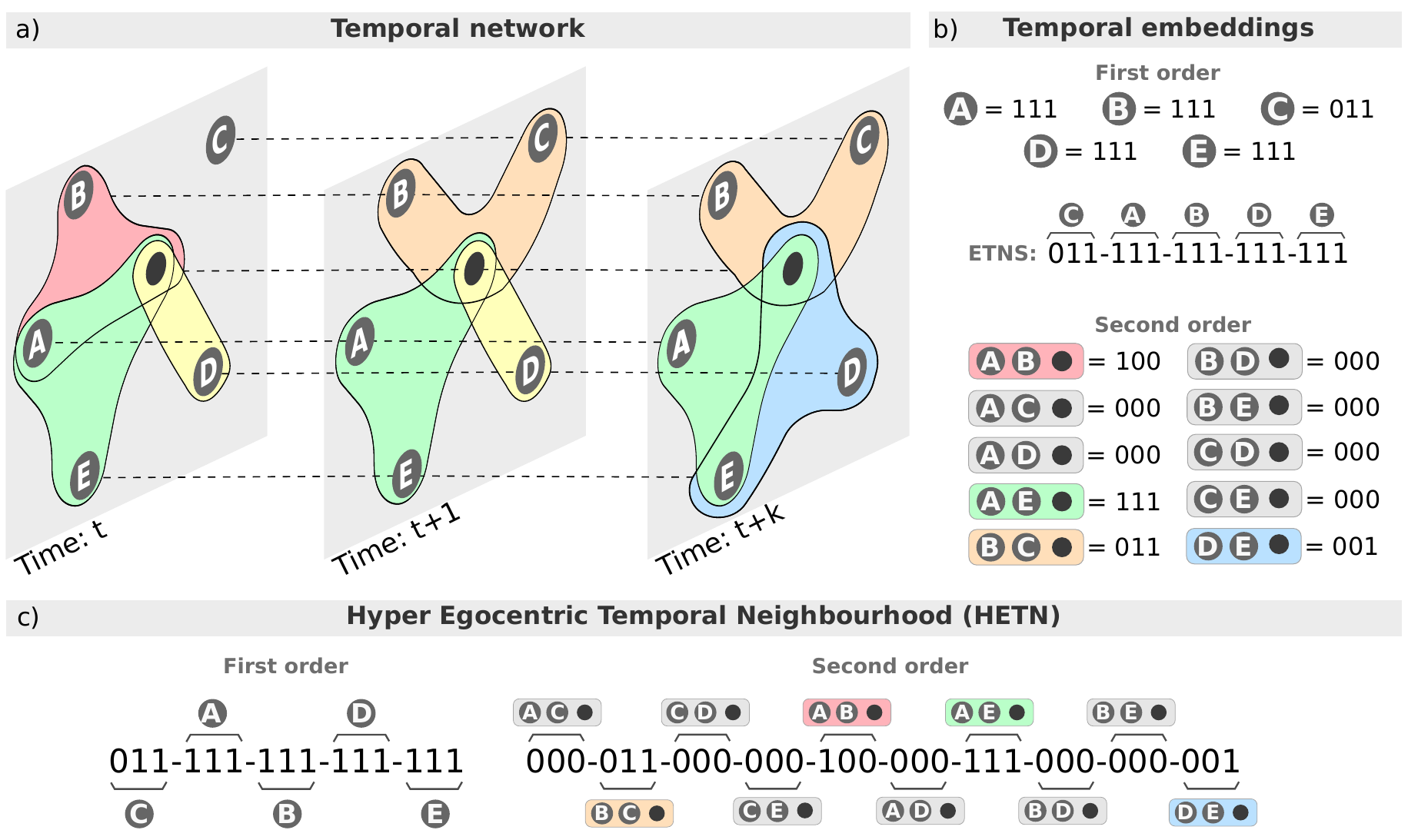}
    \caption{(panel a) Schematisation of a hyper egocentric temporal neighbourhood with $(k=2)$ and (panel b) its corresponding encoding at first and second order. (panel c) Hyper egocentric temporal neighbourhood signature (HETNS) describing the HETN in (a).
    }
    \label{fig:fig_1}
\end{figure*}

All the HETN of a temporal hypergraph can be encoded into binary signatures.
Extending the algorithm proposed in \cite{Longa2022}, the encoding at first order consists in assigning, at every time  ($t, t+1, ...,t+k$), a $1$ if the link between the node $v$ and a neighbour $j$ exists at that time or a $0$ otherwise. This results in a binary key for every node $j\neq v \in \mathcal{H}_{t}^k(v)$. The extension of the encoding to the second order is done by keeping track of the triangles (second order interactions) in which the ego node $v$ participates. A binary key is assigned for every possible dyadic interaction between nodes other than the ego node, giving a total of $n(n-1) / 2$, where $n$ is the total number of nodes in the HETN excluding the ego node. In Fig.~\ref{fig:fig_1}, for example, we can see how the key encoding the temporal interaction between the ego node and node $C$ is ``011'' as the edge $(v,C)$ does not exist at time $t$ but exists at $t+1$ and $t+2$. A similar case happens for the triangle composed of the ego node and vertices $B$ and $C$ which would be encoded by the key associated with the edge $(B,C)$ and which results in ``011''.\\

Once we have the binary keys encoding the dyadic (first order) and triangle interactions (second order), we proceed to define the key for the particular HETN by sorting lexicographically the first order encoding. The new sorting at the first order will define the sorting of the second order. For the particular case shown in Fig.~\ref{fig:fig_1} we would have: $011-111-111-111-111$ $(C - A - B - D - E)$ and consequently $100-000-000-111-011-000-000-000-000-001$ $(CA - CB - CD - CE - AB - AD - AE - BD - BE - DE)$. Finally, we will concatenate the first and second order keys: $011-111-111-111-111- -100-000-000-111-011-000-000-000-000-001$. This vector is named \textit{hyper egocentric temporal neighbourhood signature} (HETNS).

This encoding method finds its limitations when encountering isomorphic subgraphs. In fact, while the main advantage of the encoding at first order is that it bypasses the problem of isomorphic HETNs\cite{Longa2022}, this is not always possible at the second order. In fact, when there is a multiplicity of keys at the first order, several encodings at second order may be present. For instance, the example in Fig.~\ref{fig:fig_1} could also have been sorted as $C - A - B - E - D$ (and other 22 extra combinations) at first order, resulting in a different second order key for the very same HETN. Thus, in the event that there is a multiplicity of keys at the first order, every possible sorting combination at the second order must be checked in order to index isomorphic HETNs under the same, unique, HETNS. In theory, this task becomes rapidly computationally unfeasible since every element duplicated $n$ times means $n!$ possible sort combinations. However, in practice, given the sparseness of interactions of all the datasets and the aggregation times (see the datasets and hypergraphs subsection, below) considered, the number of non-computable cases is negligible. This allows us to compute the HETNS of each dataset in a few minutes.\\
\\
A binary signature allows us to store a high number of HETN. This, in turn, lets us characterise any node in a temporal hypergraph or the entire temporal hypergraph as a whole. 
Indeed, to characterise a single node $v$ according to this algorithm, it is sufficient to build the vector of frequencies of every HETN where $v$ participates as the ego node. These can be obtained for each time $t$, using a sliding window over time.
Instead, to describe a whole temporal hypergraph, we analyse the HETN of all nodes with their frequencies together. We denote these HETN-based descriptions as node (hypergraph) embedding, \textit{EMB(v)} \textit{(EMB ($\mathcal{H})$)}.\\
Using this notation, we can consider all the HETNS as the basis of a vector space and their frequencies as the coordinates. 
This allows us to compare two nodes (hypergraphs) $v_1$ and $v_2$ ($\mathcal{H}_1$ and $\mathcal{H}_2$) by computing a cosine distance between the embedding vectors as:
\begin{equation}
    \label{eq:dist}
    dist(X_1, X_2) = \left( 1- \frac{\text{EMB}(X_1) \cdot \text{EMB}(X_2)}{||\text{EMB}(X_1)|| \quad ||\text{EMB}(X_2)||} \right)
\end{equation}
where $X$ represents either a node $v$ or a hypergraph $\mathcal{H}$.\\

\subsection{Hyper egocentric temporal motifs}
The set of HETN of a hypergraph can include very basic structures with high frequency which are common to all hypergraphs, as we will see below. For some applications, it is hence useful to filter them and reduce the analysis to a smaller set of HETN that are significant to describe that specific hypergraph. This filtering procedure makes use of a null model that assumes no temporal correlations between the interactions.
We obtain a null model $\bar{\mathcal{H}}$ of $\mathcal{H}$ by randomly shuffling the  hypergraph temporal snapshots $\{ \mathcal{H}_{1}, \mathcal{H}_{2}, ..., \mathcal{H}_{T}\}$ in which $\mathcal{H}$ may be split.\\
We will consider HETNS from $\mathcal{H}$ as significant when they satisfy the following requirements: 
\begin{itemize}
    \item Over-representation: $P(\bar{N}_{\bar{\mathcal{H}}} > N_{\mathcal{H}}) < \alpha$;
    \item Minimum deviation: $N_{\mathcal{H}} - \bar{N}_{\bar{\mathcal{H}}} \geq \beta  \bar{N}_{\bar{\mathcal{H}}}$;
    \item Minimum frequency: $N_{\mathcal{H}} \geq \gamma$.
\end{itemize}

Where $N_{\mathcal{H}}$ is the number of occurrences of a HETNS in $\mathcal{H}$, $\bar{N}_{\bar{\mathcal{H}}}$ is the average number of occurrences of the same HETNS in $\bar{\mathcal{H}}$ and we set $\alpha = 0.01$, $\beta = 0.1$ and $\gamma = 5$ respectively. We refer to the resulting significant HETNS as \textit{hyper egocentric temporal motif} (HETM). The same as in the HETNS case we denote as $EMB_{HETM}(\mathcal{H})$ the embedding vector describing the graph $\mathcal{H}$ based on their HETM frequencies. Frequencies are normalized by the maximum frequency in each vector.\\

\subsection{Datasets and hypergraphs}
To validate our algorithm, we evaluate it across ten proximity contact networks. Nine of them model face-to-face interactions and have been collected using wearable proximity sensors (radio-frequency identification (RFID) tags) during the  SocioPatterns collaboration \footnote{http://www.sociopatterns.org/}. The devices record face-to-face interactions between users and data have been made available as pairwise interactions with a temporal resolution of $20$ seconds. 
Table \ref{tab:network_stat} shows the number of nodes, the number of edges and the durations (in days) for each of the following datasets: 
\begin{itemize}
    \item Three high schools, containing interactions between students and professors, gathered at a high school in Marseille (France) in the years 2011, 2012, and 2013~\cite{fournet2014contact,mastrandrea2015contact}. Metadata informs about the class of each student or of the role of the professors.
    \item  A primary school, containing interactions between students and teachers, collected in a French primary school~\cite{stehle2011high,gemmetto2014mitigation}.
    \item  Two workplaces, containing interaction between co-workers in an office building in France in 2013 and 2015~\cite{genois2015data,genois2018can}.
    \item A hospital, containing interactions between workers and patients, collected in the geriatric ward of a university hospital in Lyon (France)~\cite{vanhems2013estimating}. Metadata informs about the role of each individual (medical doctors, patients, nurses and administrative staff).
    \item A scientific conference, containing interactions between the attendees~\cite{genois2018can}.
    \item A primates dataset, containing the interactions between Guinea baboons living in an enclosure at the CNRS Primate Center in Rousset-sur-Arc (France)~\cite{gelardi2020measuring}.
\end{itemize}
In addition, we also analyse a dataset of proximity contacts collected via Bluetooth technology. Here, interactions depend only on physical distance and do not always imply an active interplay among people. This data was gathered at a university campus (at the Technical University of Denmark, DTU) where a group of freshmen students participated for a month using phones which provided physical proximity information using Bluetooth tech \cite{Sapiezynski2019}. We analyse a sample of these data comprising just one week of interactions.
\begin{figure*}
    \centering
    \includegraphics[width=0.995\textwidth]{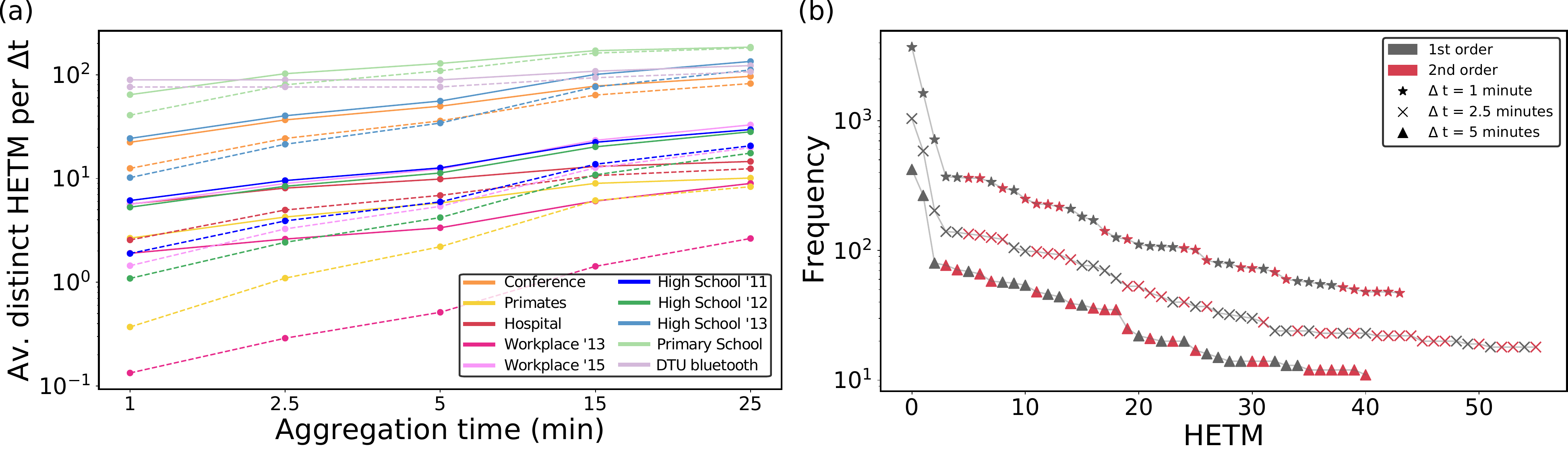}
    \caption{(panel a) Average number of distinct Hyper Egocentric Temporal Motifs (HETM) per temporal layer in proximity contact data for different aggregations of time $\Delta t$ and $k=2$. Continuous lines correspond to first and second order HETM and dashed lines to pure second order HETM. (panel b) Rank plot of the abundance of HETM in the \textit{Hospital} data (excluding the 20\%  less common HETM) for different aggregation times. Points in grey indicate the 1st order HETM and points in red are second order HETM.}
    \label{fig:fig_2}
\end{figure*}
\begin{table}[H]
    \centering
    \begin{tabular}{l|c|c|c}
         Name & $|V|$ & $|E|$ & Duration \\
         \hline
         Conference         & 403   & 70261   & 1.32   \\
         Primates           & 13    & 63095   & 28     \\
         High school 11     & 126   & 28561   & 3.1     \\
         High school 12     & 180   & 45047   & 8.44     \\
         High school 13     & 327   & 188508  & 4.2     \\
         Primary school     & 242   & 125773  & 1.35     \\
         Workplace 13       & 92    & 9827    & 11.4     \\
         Workplace 15       & 217   & 78249   & 11.5     \\
         Hospital           & 75    & 32424   & 4     \\
         DTU Bluetooth      & 656   & 717669  & 7
    \end{tabular}
    \caption{Number of nodes, number of edges and duration (in days) for each network}
    \label{tab:network_stat}
\end{table}

All these datasets come as a list of temporal pairwise interactions. Each term is encoded as $(i, j, t)$, where $i$ and $j$ represent individuals, i.e., the nodes of our hypergraphs, and $t$ is the time at which the interaction took place.
To build a hypergraph from these data, we need to choose a temporal interval $\Delta t$ that will become the time resolution of the temporal hypergraph. The timestamps appearing in the dataset are hence divided into slots of length $\Delta t$ and all the interactions that take place between $t$ and $t+\Delta t$ are stored as interactions taking place at time $t$ in the hypergraphs. We call $\Delta t$ \textit{aggregation time}. Moreover, we divide interactions into first and second order, i.e. if in an interval there are the interactions $(i,j)$, $(j,k)$ and $(k,i)$, we consider this as a unique second order hyperlink $(i,j,k)$. We encode this interaction as a hyperedge of second order, instead of three first order edges. Larger values of $\Delta t$ result in a larger number of interactions included in each temporal snapshot, thus increasing also the probability of forming second order interactions.

In the following, we will show the results obtained for several temporal hypergraphs generated from the above datasets using different aggregation times $\Delta t$.
We will start by comparing the different datasets using the distance metric defined in eq.~\ref{eq:dist} based on their $EMB_{ETM}(\mathcal{H})$. Afterwards, we will zoom in on individual nodes and compute their HETNS. We will then compute distances among them obtaining distance matrices that we represent using a multidimensional scaling technique. This is a dimensionality-reduction technique which consists of translating the values of a distance matrix into an $n$-dimension representation by preserving the distances in the $n$-dimension space. In our case, we will use an Euclidean representation. This technique allows us to visualise similarities among nodes computed from their HETNS composition while also pointing out the different roles of the nodes given by the contact metadata. We will analyse the node's HETNS nature with a focus on distinguishing the contribution provided by the pure first and pure second order HETNS. Finally, we will look at the temporal coherence of a node with itself by also looking at its HETNS embedding at different times.

\section{Results}
\begin{figure*}
    \centering
    \includegraphics[width=0.7\textwidth]{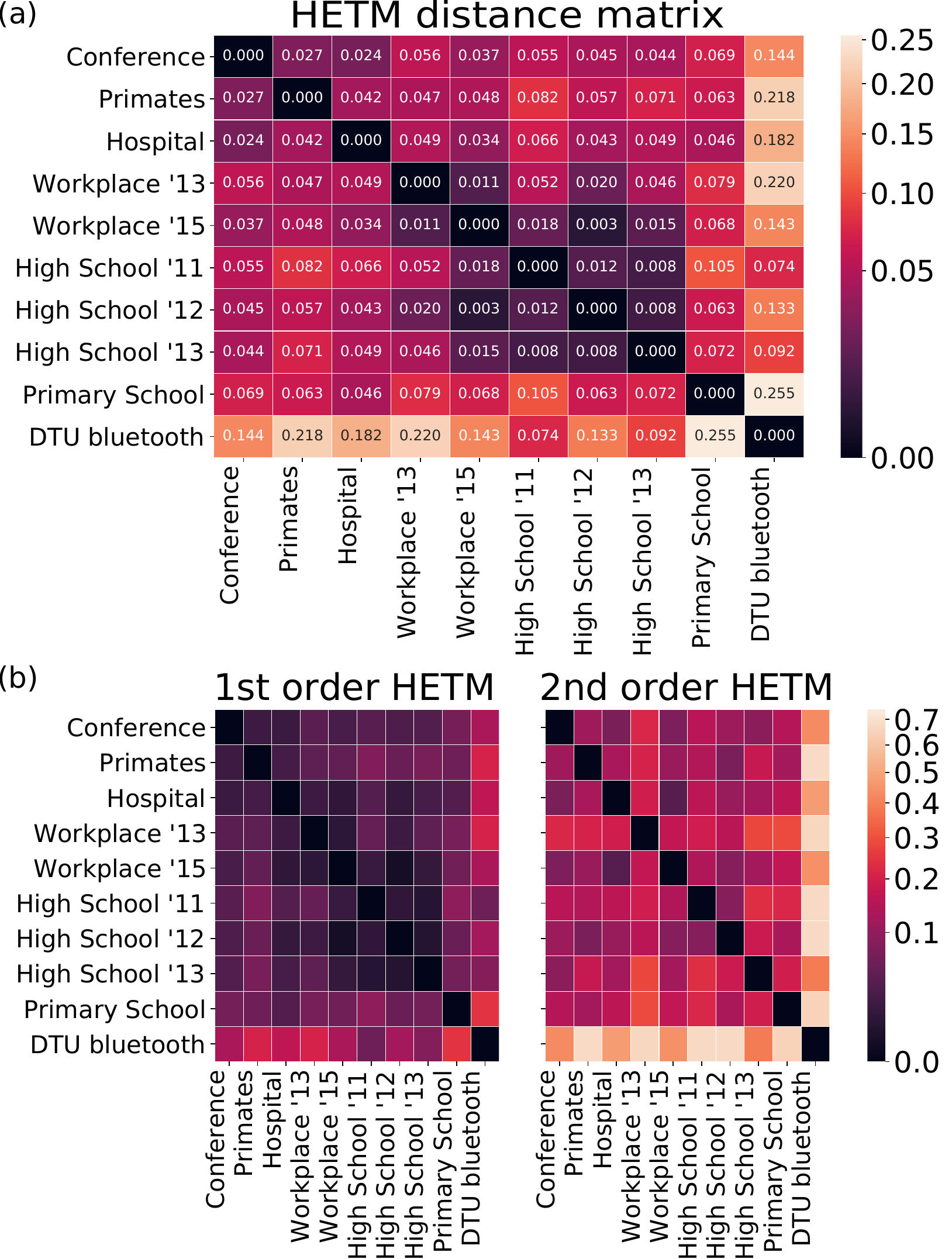}
    \caption{(panel a) Distance matrix among datasets based on HETM. Results are shown for $\Delta t = 2.5$ min and $k = 2$. (panel b) Distances computed by only considering the first or second order HETM respectively. Colour bars are represented in a logarithmic scale.}
    \label{fig:fig_3}
\end{figure*}
We start our analysis with HETM, which only include the significant structures with respect to the null model.
As expected, the number of distinct HETM in real temporal hypergraphs varies with datasets and aggregation time, as shown in Fig.~\ref{fig:fig_2} (a). We can observe that from all the distinct HETM appearing for each dataset (continuous lines), the number of second order HETM (dashed lines) is quite large, even for low aggregation times and calls for a  deeper analysis.
We observe that increasing the aggregation time makes the number of different HETM increase both for first and second order.
Considering larger aggregation times indeed implies that a larger number of nodes interact with the ego node thus increasing the complexity and hence the variability of HETM structures. 
The abundance of each HETM, i.e. the number of times that a HETM appears in the network, is reported in Fig.~\ref{fig:fig_2} (b) for the \textit{Hospital} dataset. Again, although, for each aggregation time analysed, the first four most frequent HETM only involve pairwise interactions, starting from the fifth first and second order HETM seem to be equally likely. 
Finally, in the following analyses, we will choose a small aggregation time (2.5 minutes) because we recall that choosing a value of $\Delta t$ implies considering interactions taking place in that range as simultaneous, which can be considered a good approximation only for ranges of few minutes.

The set of HETM obtained from a temporal hypergraph schematises the behaviour of the nodes of that specific hypergraph with respect to their neighbourhood. Then, they can be used to characterise and compare different datasets. 
To do so, we define the distance between two datasets as the inverse of the cosine similarity between the two corresponding vectors of HETM frequency embeddings (see definition in Methods). Fig.~\ref{fig:fig_3} shows the matrix of pairwise distances between all the considered datasets. The top panel reports distances computed considering all the HETM, while the bottom panels have been obtained by only using purely first and purely second order respectively. 
From the top panel, we notice a higher similarity between high schools and between workplaces. The Bluetooth dataset instead appears more distant from all the other ones, probably because of the different technology that implies a larger variety of interactions. Bluetooth data, in fact, records co-location between individuals, not only face-to-face, so a recorded connection between two nodes means that the corresponding individuals are in the same area, not necessarily interacting. This results in larger groups and hence the appearance of more complex HETN than for the other datasets.
Interestingly, distances at the first order are smaller with respect to considering both types of interactions, suggesting that first order HETM are very similar between datasets. On the contrary, the second order one shows generally larger distances among datasets and accentuates the distance of the Bluetooth dataset from the other ones.
\begin{figure*}
    \centering
    \includegraphics[width=0.95\textwidth]{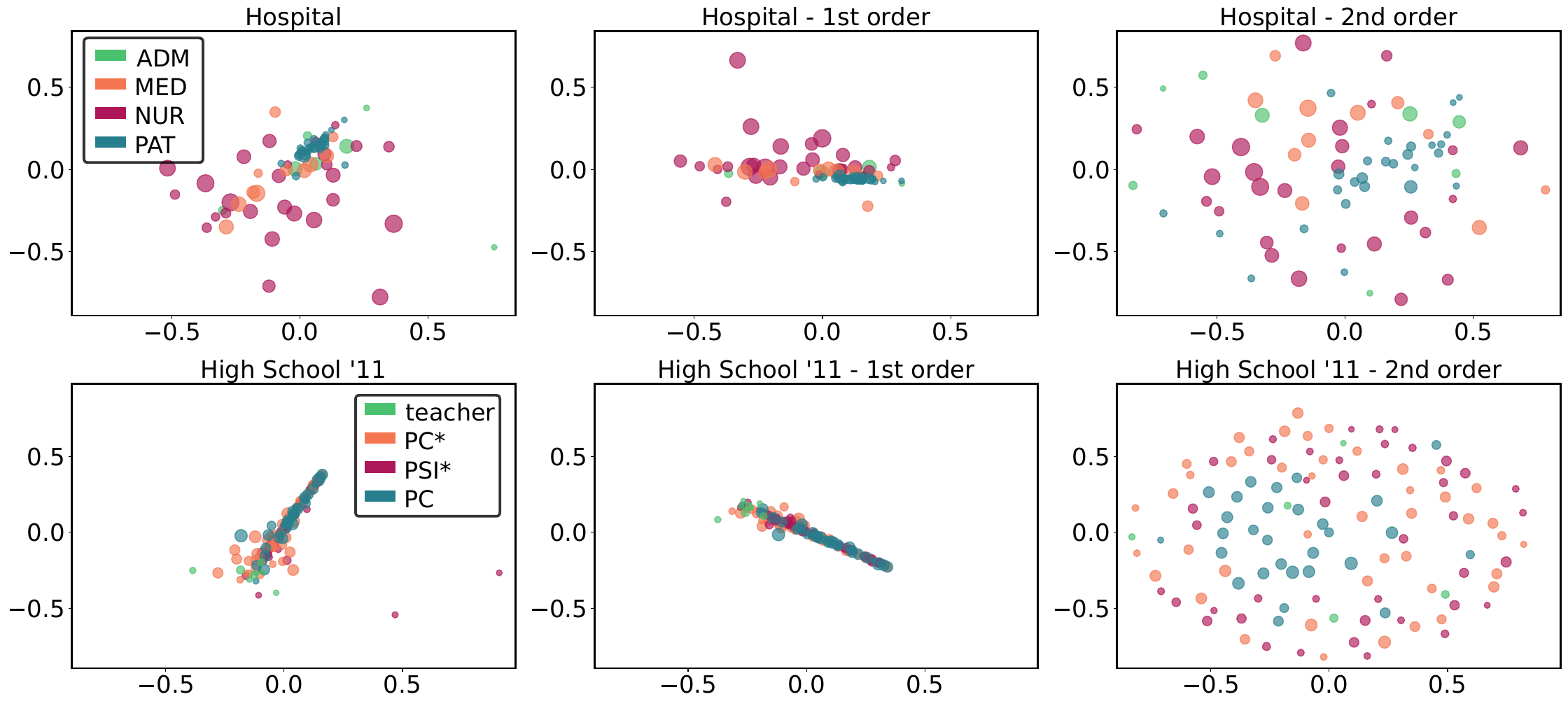}
    \caption{Multidimensional scaling of HETN-based distances at single node level with $\Delta t = 2.5$ min $k=2$ for two different proximity contact datasets: \textit{Hospital} (upper panels) and \textit{High School'11} (lower panels). We show the multidimensional scaling accounting for all the HETN (left panels) and those obtained considering only first (central panels) and second (right panels) order HETN. The size of the nodes is proportional to the number of different HETN that each node presents. In the \textit{Hospital} dataset (upper panels) the colour of the nodes corresponds to four different classes: administrative staff (green), medical doctor (orange), paramedical staff (purple) or patient (blue). In the \textit{High School'11} dataset (lower panels) the colour of the nodes depends on whether the node is a teacher (green) or a student belonging to three different classes: PC* (orange), PSI* (purple), PC (blue).}
    \label{fig:fig_4}
\end{figure*}

\begin{figure*}
    \centering
    \includegraphics[width=0.975\textwidth]{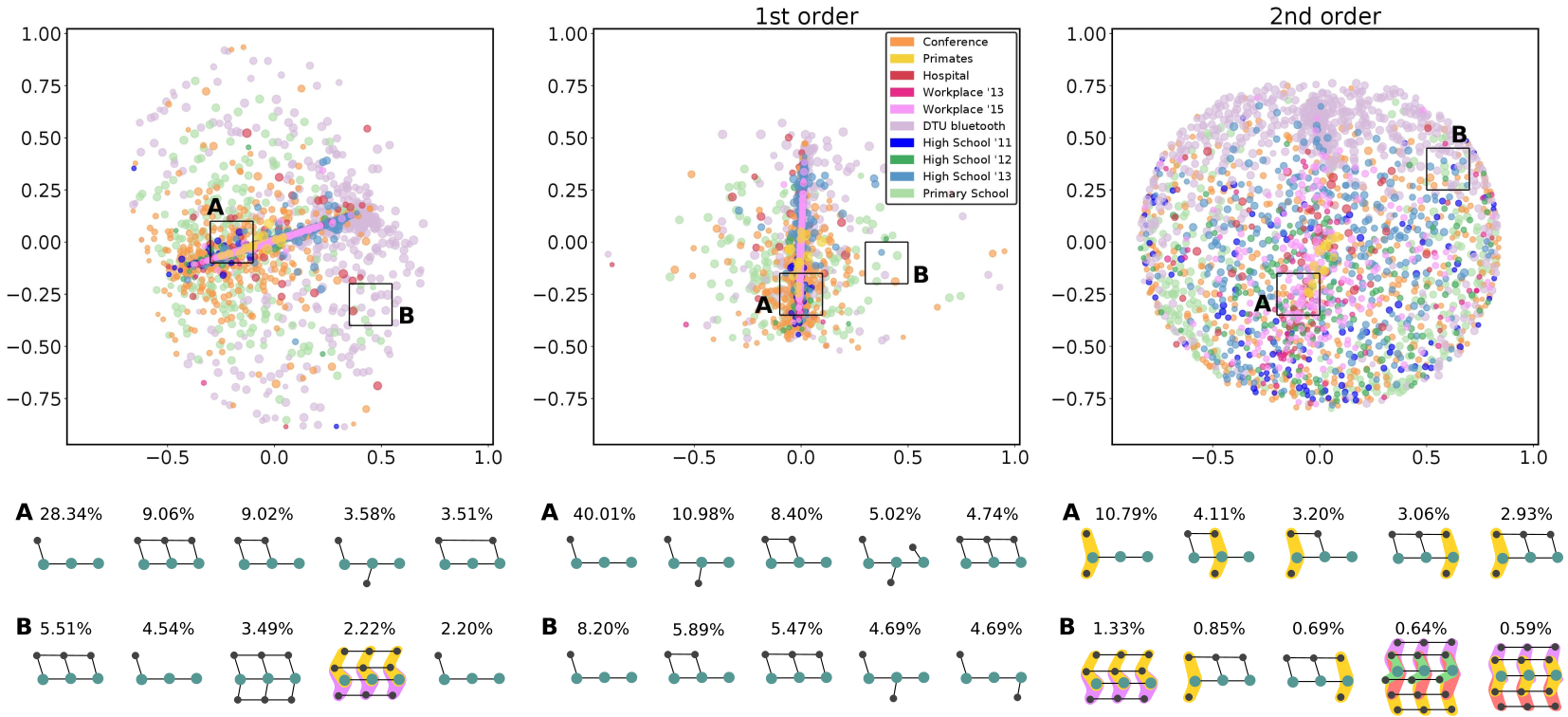}
    \caption{(upper panels) Multidimensional scaling of HETN-based distances at single node level for $\Delta t = 2.5$ min and $k=2$ for the 10 different social contact data considering both (left), first (middle) and second (right) order HETN. (lower panels) each row shows the five most frequent HETN in their corresponding patch displayed in the cartography above. The percentage upon each HETN indicates the percentage in which that signature appears among all the HETN of the nodes in the patch.}
    \label{fig:fig_5}
\end{figure*}

If the hyper egocentric temporal neighbourhoods observed in one network characterise the corresponding dataset, those found for one node characterise the behaviour of the corresponding individual. We hence use them to compare individuals and identify classes of behaviours. 
By using a distance metric analogous to the one used above, we implement a multidimensional scaling, i.e. we position nodes in a 2D-space according to the reciprocal distance (see Methods).  Notice that for nodes' analysis (figs.~\ref{fig:fig_4} and \ref{fig:fig_5}) we make use of the entire set of HETN, not only the HETM. In fact, the concept of HETM has been defined to characterise a hypergraph and not singular nodes; the structures that are significant for a hypergraph do not necessarily coincide with the structures that are significant for a node. Since the concept of significance for a node (or a group of nodes) is not a trivial matter, we choose to use all the collected HETN to characterise a node.

In Fig.~\ref{fig:fig_4} we report for two datasets (\textit{High School'11} and \textit{Hospital}) the multidimensional scaling of nodes' distances. The size of the nodes provides information on the number of different HETN: larger nodes have a larger variety of HETN. Besides, each node may belong to a different class (information provided by metadata) which is represented by its colour.
For both datasets 3 panels are shown: the distances obtained comparing all HETNS, those obtained from the first order only, and those from the second order only, respectively.
From the comparison between all HETNS (1st and 2nd order together) for the \textit{High School'11} dataset (lower panels of Fig.~\ref{fig:fig_4}), we notice a central cloud containing $98\%$ of the nodes. In this cloud, we see that nodes representing students in different classes (PC*, PSI* and PC) are well mixed, while ones corresponding to Teachers are located on the edge of the cloud. Besides, this last group has, on average, a smaller size, meaning that they present a lower variety of behaviours.
For the \textit{Hospital} (upper panels of Fig.~\ref{fig:fig_4}) we observe a higher dispersion, but we can easily appreciate that the separation between different classes of nodes partially reflects the metadata. Indeed, we notice a small cluster corresponding to the patients, which is also the class showing the lowest number of different HETN. This cluster is surrounded by that of the administrative staff and another, more dispersed, composed of medics. Nurses are instead spread around, suggesting a more varied behaviour, not characterised by a small group of specific HETN as for the other classes.

Again, when we restrict the analysis to the first order, we observe in both cases a behaviour similar to the previous one, with a similar clustering of the nodes. The main difference is represented by the fact that the clouds are smaller, implying a general higher similarity between individuals. 
On the other hand, the second order panels exhibit more dispersed clouds, with distances approximately twice those computed for the first order. This means that the second order reveals a higher variability of temporal neighbourhoods among different nodes, something that could not be appreciated if limiting the analysis to the first order.

After analyzing individual HETNS within each dataset, it might also be instructive to investigate whether classes of behaviors appear consistently across datasets. 
To do that, in Fig.~\ref{fig:fig_5} we report the multidimensional scaling for all the nodes of all the analysed datasets, where different node colours identify different datasets. Again, we show the results for all HETN and those for the first order only and the second order only.

By analysing first and second order together we notice that most of the nodes are concentrated in the central part of the plane --i.e. many HETNS are shared among datasets. There, we can find the smaller and simpler structures, involving mainly the first order, (as can be seen by the five most frequent ones collected in region $A$ and reported on the bottom). These are common to all datasets.
Moving away from that central region more complex structures start to appear, involving second order too. Here the datasets start to differentiate: the two workplaces rapidly disappear (implying that only very basic structures are present in these datasets); some other datasets (\textit{Primates}, \textit{Conference}, \textit{Hospital}, and the high schools) show only a few individuals that behave differently than the majority; while \textit{Primary school} and the \textit{DTU Bluetooth} are very spread in space, with a large variety of behaviours. 
In the peripheral region $B$ we notice that second order structures,  quite uncommon in region $A$, are not only present but significant, also appearing among the 5 more frequent HETN.

The central panel of Fig.~\ref{fig:fig_5}, depicting the results limited to the first order, shows a different pattern, where most of the datasets are collapsed in the middle and only a few nodes differentiate. Here the two regions $A$ and $B$, central and peripheral, show very similar and simple structures(four HETN out of five are common to the two areas). Moreover, in both cases, the five reported structures cover more than 25\% of the structures (60\% for the central region). The variability between possible structures that we find at the first order is hence very limited.

In contrast, the second order structures (right panel) compose a larger cloud where all nodes are more spread around, revealing a larger variability of second order HETN for nodes of all datasets (except for the primates one).
Here there is not a particular region where the concentration of points is denser than in the others. This reveals that there are not many structures that are common to different nodes, as confirmed by observing the reported structures below: the most frequent HETN on each region covers less than the 25\% of the total (less than the 5\% for the outer region).

\begin{figure}[H]
    \centering
    \includegraphics[width=0.495\textwidth]{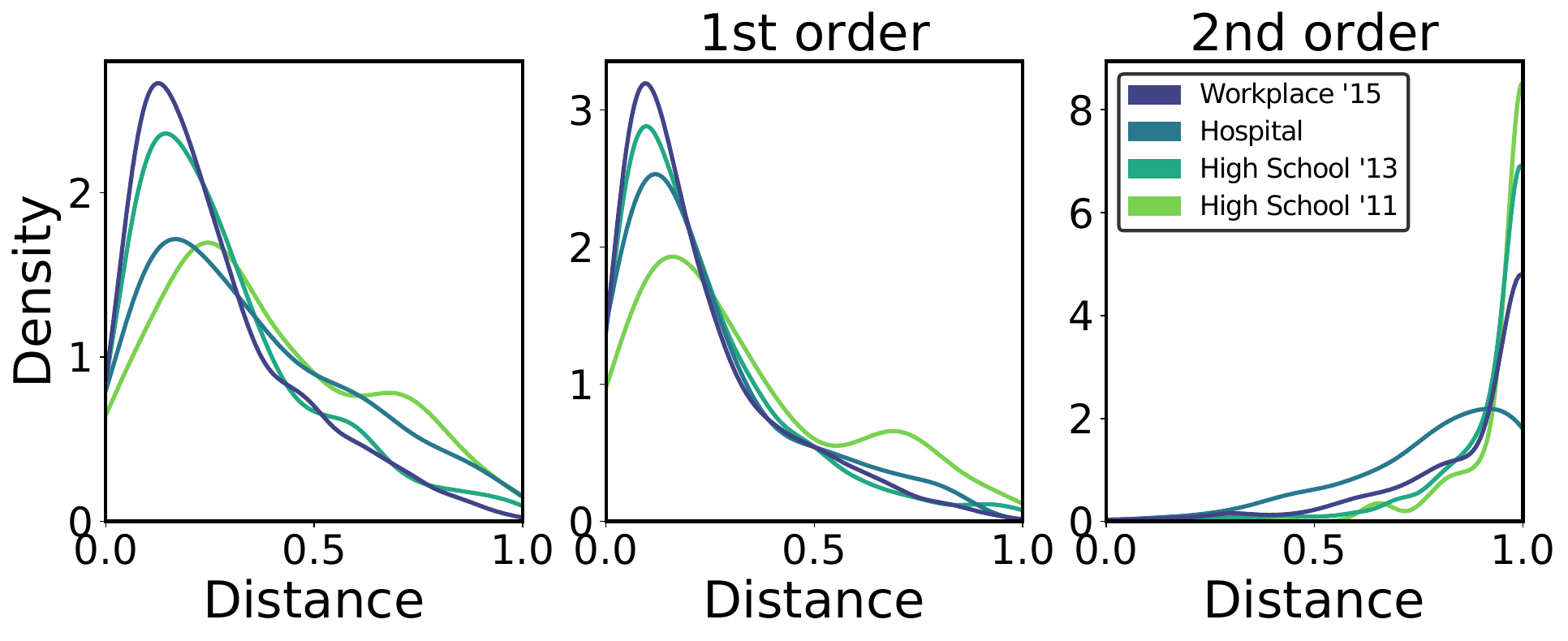}
    \caption{Temporal distance distributions computed for every node with itself comparing their HETN embeddings  for $\Delta t = 2.5 $ min and $k=2$. Distances are calculated evaluating the HETN embedding of each node in consecutive daily time windows for each dataset. Each \textit{pdf} is normalised so that the total area of each histogram equals one.} 
    \label{fig:fig_6}
\end{figure}

Finally, once analysed HETNS variability among datasets and between nodes, we can also study how individuals change their behaviour over time. 
In Fig.~\ref{fig:fig_6}, instead of comparing different nodes, we compare each node with itself at different times. In particular, we divide the time series of interactions in daily intervals ($[T_0, T_1, ..., T_N]$). The HETN extracted for one node at $T_0$ are hence compared with those extracted for the same node in $T_1$; the same for $T_1$-$T_2$ up to $T_{N-1}$-$T_N$. The panels report the distribution of distances computed for all the nodes considering all orders together, the purely first order and the purely second, respectively. The difference between the overall behaviour, first and second order, is even greater, with distances that pass from an average of around 0.24 for first order only to an average of around 0.90 for second order.

Taken together our results show that a high variability between second order structures (both among different nodes and among different temporal intervals of the same node) is always observed.
This supports the idea that a higher-order analysis is hence needed to distinguish the social behaviours of different nodes and in different datasets. Focusing only only first order structures is indeed limiting because most of the extracted HETN are trivial and common to all nodes and to all datasets, not allowing to reveal singular peculiarities. To clean the data from these trivial structures requires the use of a null model (as we did in the HETM analysis) which can however introduce biases. The second order structures instead allow more refined analyses. 

\section{Conclusions}
Considering the temporal dimension is fundamental to understand social interactions and, in this sense, temporal networks provide a unique tool for their analysis. Nevertheless, capturing and modelling all the complexity of temporal interactions is an open question in network science.

To represent these temporal structures,  egocentric temporal neighbourhoods/motifs have emerged as a concise yet powerful tool to analyse and model their dynamics. Focusing only on ego nodes and their connections allows us to analyse each node individually and represent the evolution of the network in a parsimonious way. In the last years, however, a new paradigm emerged highlighting the role of higher-order structures in shaping the dynamics of social systems. In this approach, the emphasis is on correlations in the interactions between neighbours of the ego nodes: something that the original definition of temporal neighbourhood neglects by construction~\cite{Longa2022}. Thus, the question of how relevant are higher-order interactions for the modelling of this class of dynamics naturally arose.

To answer this question, in this work, we used a generalisation of temporal graphs to temporal hypergraphs and we extended the concept of egocentric temporal neighbourhood to include second-order interactions -- connections involving two neighbours of the ego node that are also interacting between them -- such defining hyper egocentric temporal neighbourhood.  This extension allowed us to analyse several datasets from face-to-face interactions covering different contexts and aggregation times. Datasets range from interactions between students in primary and high schools to workplaces and hospitals and baboons living in a research facility. 

While first-order signatures are more abundant, as expected, second order ones still play an important role. In fact, they are present in the most frequent structures in all the datasets, with their signatures being statistically relevant even for larger aggregation times. 

Interestingly, we also find that, at the whole hypergraph level, first order structures are way more similar than second order ones. With the latter showing a larger variability and diversity. This translates into most of the differences between hypergraphs being due to second order HETM.

This same result also holds if we zoom in at the individual nodes level: second order structures are more sensitive with respect to nodes' dissimilarities. Moreover, inside each dataset, these distances between nodes' signatures allow us to better cluster individuals according to their function --i.e. nodes with similar functions such as patients or administrative staff in the \textit{Hospital} dataset, show similar signatures. 

Finally, looking at distances between signatures of the same nodes at different periods in time, we demonstrate that second order structures also show a higher temporal variability.  

Taken together, our results demonstrate that, while first order HETM provide a backbone of social interactions, second order ones are fundamental for capturing their heterogeneities at all scales: the whole network, between nodes, and temporally, even at the individual level. This picture demonstrates, once more, the relevance of higher-order structures in shaping social and network dynamics in general.

However, the extension of egocentric temporal neighbourhoods to second-order motifs comes at a cost. Second order encodings, unlike first-order ones, are not unique, forcing us to test for isomorphic patterns; a major disadvantage that first-order analysis overcame. However, this limitation is mainly theoretical. For all the datasets and aggregation times considered, the computational costs of the isomorphism tests are almost negligible, with a running time for computing nodes' signatures of a few minutes for the largest dataset and longest aggregation time. A second limitation of our work is the fact that we limit our analyses to second order motifs instead of considering third or even higher-order structures. However, their limited number suggests that most of the connections of the networks would be covered by considering first and second order HETM only.

In conclusion, with this work, we demonstrate that considering hyper egocentric temporal motifs in the analysis of temporal datasets is fundamental to capture the intrinsic variability of social interactions. In this sense, we pave the way for a deeper understanding of temporal social interactions with repercussions in their modelling and generation of synthetic surrogates.

\begin{acknowledgments}
B.A.G. is partially supported by Mar\'ia de Maeztu Program for Centers and Units of Excellence in R\&D, grant MDM-2017-0711 funded by MCIN/AEI/10.13039/501100011033. B.A.G. and S.M. acknowledge funding from the  APASOS project  PID2021-122256NB-C22 funded MCIN/AEI/10.13039/501100011033 / FEDER, UE (Away to make Europe) and  by the Mar\'ia de Maeztu Program for Centers and Units of Excellence in R\&D, grant CEX2021-001164-M funded by the  MCIN/AEI/10.13039/501100011033.
A.L. acknowledges the support of the MUR PNRR project FAIR - Future AI Research (PE00000013) funded by the NextGenerationEU.
G.C. acknowledges the support of the European Union’s Horizon 2020 research and innovation program under the Marie Skłodowska-Curie grant agreement No 101103026.\\
\end{acknowledgments}

\bibliography{ref}



\end{document}